\documentstyle[prb,preprint,aps]{revtex}

\begin{document}

\title{Hall effect and geometric phases in Josephson junction arrays }

\author{P. Ao$^1$, Y. Tan$^2$, and X.-M. Zhu$^1$}

\address{ {$^1$Department of Theoretical Physics ,
Ume\aa{\ }University, S-901 87, Ume\aa, SWEDEN } \\
{$^2$Department of Physics and Applied Physics,
 University of Strathclyde, Glasgow G4 0NG, UK}   }

\maketitle 

\abstract{ Since effectively the local contact 
vortex velocity dependent part of the Magnus force in
a Josephson junction array is zero in the classical limit, we predict
zero classical Hall effect.
In the quantum limit because of the geometric phases due to 
the finite superfluid density at superconductor grains, 
rich and complex Hall effect is found  
in this quantum regime due to the Thouless-Kohmoto-Nightingale-den-Nijs
effect.\footnote{This work was supported by Swedish NFR. } }



Since vortex dynamics is identical to that of an electron in the 
presence of a magnetic field, numerous models
for the quantum Hall effect in both homogeneous and inhomogeneous
superconductor films have been proposed.\cite{quantumhall}
\footnote{ to appear in Czechoslovak J. Phys., Vol.46, Suppl. S2, pp589 (1996)}
Those models have fully explored the analogy to the 
quantum Hall effect in semiconductor heteorojunctions by treating vortices
moving in a uniform magnetic field with a homogeneous background. 
While the treatment can be justified in a homogeneous superconductor film, 
it may not be so in inhomogeneous cases such as in Josephson junction arrays,
where both the (fraction part) number of magnetic flux, the frustration $n$,
and the (fraction part) of the fictitious magnetic flux $\phi_0$
per plaqutte are usually large, and 
the periodic potential for a vortex is strong.
Therefore the simple contineous limit without the accounting of 
the potential is not appropriate for a Josephson junction array. 

In a Josephson junction array, because of the huge core energy, 
vortices cannot move into the superconducting gains. 
They are confined to move along the junctions and the 
voids (nonsuperconducting areas), an example of the guided vortex motion.
Since the vortex velocit part of the Magnus force is proportional to
the local superfluid density, derivable
from the nonlinear Schr\"{o}dinger Lagrangian formulation\cite{zhu}, 
this force is zero for a vortex at a void, and, exponentially small
at a junction.
Furthermore, because of the guide motion, even the small transverse 
at a junction does
not produce a sideway motion.
This implies that this 
local contact transverse force does not play a role in vortex dynamics
in a Josephson junction array. Hence there is no Hall effect 
in the classical limit. 
This absence of the {\it en route}
transverse force is in agreement with  experimental 
observations\cite{experiment}. 
The condition for the classical limit will be given at the end of the paper.

In the quantum regime, however, vortices experience geometric phases similar
to the Aharonov-Bohm effect, 
due to the finite superfluid density at superconductor grains.\cite{ao}
To be commensurated with the existence of the vortex inaccessible
regions and the geometric phases,
we consider the the tight-binding limit of vortex motion. 
The corresponding Hamiltonian may be written as
\begin{equation}
   H= t \; \sum_{(l,j)} a^{\dag}_l a_j \; e^{i A_{lj}}
        + \sum_{l,j} a^{\dag}_l a_l \; V_{lj} \; a^{\dag}_j a_j \; ,
\end{equation}
where $ a_l $ is the boson annihilation 
operator for a vortex at j-th void, and 
$( \; )$ stands for the summation over nearest neighbors.
The phase $A_{lj}$ is
defined on the links connected the nearest neighbors, and its 
sum around a plaquette is equal to the geometric phase $2 \pi \phi_0$:
$\sum_{plaquette} A_{lj} = 2 \pi \phi_0$. A uniform
geometric phase in a square lattice 
will be assumed, where the number of `fluxes'  $\phi_0$ 
is the number of Cooper pairs on a superconductor grain, which 
may be controlled by a gate voltage.
The interaction between vortices is described by $V_{lj}$, which is long range 
and repulsive. We will treat it as a short range repulsive interaction for
a first approximation, further approximated by the hard-core conditions. 
The tunneling matrix element $t$ is, 
\begin{equation}
  t \simeq \sqrt{E_J E_C} \; \exp\{ - O(1) \sqrt{E_J/E_C} \} \; , 
\end{equation}
where $E_J$ is the Josephson junction energy and $E_C$ the junction charging 
energy\cite{theor}. 

To discuss the Hall effect of the idealized vortex
problem in the quantum regime, we map the hard-core boson problem onto a 
fermion problem by attaching odd number of `fluxes' on 
each vortex.
The resulting Hamiltonian for the fermion problem is 
\begin{equation}
   H= t \; \sum_{(l,j)} c^{\dag}_l c_j \; e^{i[A_{lj} + {\cal A}_{lj} ]} \; , 
\end{equation}
where $c_j$ is the corresponding the fermion annihilation operator at the j-th
void.
The number of statistical fluxes $\phi_s$ 
at the j-th void satisfys the constrain
$  \phi_s = - (2m + 1)  < c^{\dag}_j c_j > $, with 
$\sum_{plaquette} {\cal A}_{lj} = 2\pi \phi_s$, 
which means that $2m + 1$ 
fluxes have been attached to each vortex.\cite{fradkin}
If this mapping gives a mean field solution with an energy
gap separated from its excitations, 
the statistical fluxes can be adiabatically smeared over the lattice and
effectively detached from vortices.
In this case $\phi_s = - (2m + 1) n $, with $n$ is the magnetic flux 
frustration, the number of vortices per plaquette. 
Then the  resulting mean field problem is exactly 
the Harper-Azbel-Wannier-Hofstadter problem, where energy gaps do exist.
The quantum Hall behaviors of such a problem have been studied in detail
by Thouless, Kohmoto, Nightingale, and den Nijs\cite{thouless}.
For such a system the quantum Hall conductance $\sigma^f_H$ is
$
   \sigma^f_H = t_r \; ,
$
with the integer $t_r$ the solution of the Diophantine equation
$
   r = s_r q + t_r p \; .
$
Here the number of fluxes per plaquette $\phi = \phi_0 - \phi_s = p/q$, 
with $p$ and $q$ coprime,
$n = r/q$, and $r$, $s_r$, $t_r$ integers with $|t_r| \leq q/2$.
Counting the mapping generated Chern-Simons contribution to the 
Hall conductance,
$   \sigma^s_H = \frac{1}{ 2m + 1 }  ,
$
the Hall conductance of the original vortex system is
then\cite{fradkin,read}
$ 
  {1}/{\sigma^v_H } = {1}/{\sigma^f_H } + {1}/{\sigma^s_H } \; .
$
Converting back into the electric Hall conductance and putting back the unit,
we find the electric quantum conductance of the Josephson junction array is
\begin{equation}
   \sigma_H = \frac{4e^2} {h} \; 
       \frac{\sigma^f_H + \sigma^s_H }{\sigma^f_H \sigma^s_H } \;.
\end{equation}
As known in the previous study of quantum Hall effect\cite{read,fradkin}
for a given set of the `flux' $\phi_0$ and the frustration $n$, 
there may exist several values of $m$, that is, several mappings,
with their mean-field solutions all corresponding to filled bands which are
separated from excitations by energy gaps.
If such a case occurs, detailed calculation is needed fo find the $m$ with
the largest energy gap, which is the most stable one.

One can check that following symmetries hold for the quantum Hall conductance
$\sigma_H$:  the periodicity,
$\sigma_H(\phi_0, n) =   \sigma_H(\phi_0 + 1, n)$ ;
the odd symmetry, $\sigma_H(\phi_0, n) = - \sigma_H(-\phi_0, n)$; 
the particle-hole symmetry, $\sigma_H(\phi_0, n) = - \sigma_H(\phi_0, 1 - n)$.
%
%
We note that  both positive and negative Hall conductance may be easily 
reached, 
contrast to the previous proposal of the quantum 
Hall effect in a Josephson junction array\cite{quantumhall}.
For the special mapping 
$2m \; n = \phi_0 $
 the mean-field solution is automatically within a gap, and the
Hall conductance is
$   \sigma_H = 2m \; \frac{4 e^2 }{h}  .
$
With these specific sets of $\phi_0$ and $n$ and in the zero limit 
of their fraction parts 
one can take the continuous limit of the tight-binding 
model. 

 
We conclude by discussing of a criterion for the classical limit,
in which there is no Hall effect.
The relevant energy scale is the tunneling matrix element $t$.
When the temperature is higher than $t$, thermal fluctuation will destroy the
quantum coherence and the vortices move classically.
The quantum regime is realized for temperatures lower than $t$ where 
the phase coherence is preserved. 
For a Josephson junction energy $E_J \sim 1$ K and the junction charging 
energy $E_C \sim E_J $, $t \sim 100$ mK.
We point out that the Hall effect in the quantum regime may have been realized 
experimentally\cite{chen}.

\end{document}